\documentclass[traditabstract]{aa}
\usepackage{graphicx}
\usepackage{txfonts}
\usepackage{todonotes}
\usepackage[normalem]{ulem}
\usepackage{amstext}
\usepackage[breaklinks=true]{hyperref}
\usepackage{xspace}
\usepackage{lscape}

\usepackage{listings}
\makeatletter
\renewcommand*\maketitle{%
  \thispagestyle{firstpage}
\begingroup
    \if@wideboxfn
    \setlength\bibindent{1.4\parindent}
    \else
    \setlength\bibindent{\parindent}
    \fi
    \renewcommand*\thefootnote{\@fnsymbol\c@footnote}%
    \renewcommand\@makefntext[1]{%
    \ifaa@longfn\hsize\textwidth\fi
    \noindent
    \hb@xt@\bibindent{\hss\@makefnmark\enspace}##1}
  \ifaa@twocolumn
  \begingroup
    \begin{aa@strip}
          \aa@maketitle
    \end{aa@strip}
    \@thanks            
  \endgroup
  \else
    \begingroup
      \let\thanks\footnote
      \aa@maketitle
    \endgroup
  \fi
\endgroup
  \setcounter{footnote}{0}%
}
\makeatother

\definecolor{dkgreen}{rgb}{0,0.6,0}
\definecolor{gray}{rgb}{0.5,0.5,0.5}
\definecolor{mauve}{rgb}{0.58,0,0.82}
\newcommand{\orcit}[1]{\protect\href{https://orcid.org/#1}{\protect\includegraphics[width=8pt]{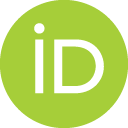}}}

\newcommand{\Gaia}{\textit{Gaia}\xspace}

\def\ltsima{$\, \buildrel < \over \sim \,$}
\def\simlt{\lower.5ex\hbox{\ltsima}}
\def\gtsima{$\, \buildrel > \over \sim \,$}
\def\simgt{\lower.5ex\hbox{\gtsima}}

\begin{document}

\title{Metal-poor stars with disc-like orbits
} 
\subtitle{Traces of the Galactic Disc at very early epochs?}
\authorrunning{M. Bellazzini et al.}
\titlerunning{Metal-poor stars with disc-like orbits}

\author{
M.~                    Bellazzini\orcit{0000-0001-8200-810X}\inst{1}
\and         D.~                       Massari\orcit{0000-0001-8892-4301}\inst{1}
\and         E.~                   Ceccarelli\inst{3,1}
\and         A.~                     Mucciarelli\orcit{0000-0001-9158-8580}\inst{3,1}
\and         A.~                     Bragaglia\orcit{0000-0002-0338-7883}\inst{1}
\and         M.~                        Riello\orcit{0000-0002-3134-0935}\inst{2}
\and         F.~                     De Angeli\orcit{0000-0003-1879-0488}\inst{2}
\and         P.~                   Montegriffo\orcit{0000-0001-5013-5948}\inst{1}
}
\institute{
INAF - Osservatorio di Astrofisica e Scienza dello Spazio di Bologna, via Piero Gobetti 93/3, 40129 Bologna, Italy
\and 
Institute of Astronomy, University of Cambridge, Madingley Road, Cambridge CB3 0HA, United Kingdom
\and 
Dipartimento di Fisica e Astronomia, Universit\`a degli Studi di Bologna, Via Piero Gobetti 93/2, 40129 Bologna, Italy$\relax                                                                                                                                                                                                                                                      $\label{inst:0003}\vfill
}



\date{Accepted for publication by Astronomy \& Astrophysics}

\abstract{We use photometric metallicity estimates for about 700000 stars in the surroundings of the Sun, with very accurate distances and 3-D motions measures from Gaia DR3, to explore the properties of the metal-poor ($-2.0<$[Fe/H]$\le -1.5$; MP) and very metal-poor ([Fe/H]$\le -2.0$; VMP) stars with disc kinematics in the sample. 
We confirm the presence of a significant fraction of MP and VMP stars with disc-like orbits and that prograde orbits are prevalent among them, with prograde to retrograde ratio $P/R\sim 3$.
We highlight for the first time a statistically significant difference in the distribution of the Z-component of the angular momentum ($L_Z$) and orbital eccentricity between prograde and retrograde disc-like MP stars. The same kind of difference is found also in the VMP subsample, albeit at a much lower level of statistical significance, likely due to the small sample size. We show that prograde disc-like MP and VMP stars display an additional component of the $|L_Z|$ distribution with respect to their retrograde counterpart. This component is at higher $|L_Z|$ with respect to the main peak of the distribution, possibly hinting at the presence of a pristine 
prograde disc in the Milky Way. This hypothesis is supported by the results of the analysis of a large sub-sample dominated by stars born in-situ. Also in this case the prevalence of prograde stars is clearly detected at [Fe/H]$\le -1.5$ and their $|L_Z|$ distribution is more skewed toward high $|L_Z|$ values than their retrograde counterpart. This suggests that
the seed of what will eventually evolve into the main disc components of the Milky Way may have been already in place in the earliest phases of the Galaxy assembly.}


\maketitle

\section{Introduction}
\label{sec:introduction}

Metal deficient stars are expected to bear memory of the earliest phases of Galaxy evolution \citep[see, e.g.,][]{miranda16,elbadry18}. 
Once believed to live only in the pressure-supported Galactic spheroid, it is now generally recognised that stars with [Fe/H]$<-1.0$ as well as with [Fe/H]$<-2.0$ exist in significant number also on disc-like orbits, either as the metal-weak component of the thick disc \citep[][and references therein]{ruchti10,beers2014,li_zhao17,carollo19} or as separate components, possibly of accretion origin \citep[see, e.g.,][and references therein]{carter21,sotillo23}.

The interest on these ancient disc components has been recently revived by the discovery of extremely metal-poor stars ([Fe/H]$\le -2.5/-3.0$) with disc-like kinematics and a preference for prograde over retrograde orbits \citep{sestito19,sestito20,cordoni21,carter21,carollo23}. The comparison with simulations of Milky Way (MW) analogues suggests that the majority of these stars should originate from accretion events on nearly planar orbits but in-situ formation may also play a significant role, depending on the specific evolutionary path of each given galaxy and on the considered metallicity range,
as well as on the specific set of simulations considered \citep{sestito21,carter21,santistevan21,carollo23,sotillo23,pinna23}. For instance, according to \citet{santistevan21} and \citet{sestito21}, the contribution of the in-situ component is minor or negligible, respectively.
In any case, these stars can provide a precious view on the earliest phases of the formation of the MW disc \citep{aurora22,conroy22,rix22,chandra23,semenov23a, semenov23b}.

In the present contribution we attempt to have a new insight into the properties of the metal-poor stars with disc-like orbits by using the recently released sample by \citet[][hereafter B23]{B23}, that provides photometric metallicity with typical accuracy of $\simeq 0.1$~dex and typical precision $\la 0.2$~dex for about 700000 red giant branch stars 
within a few kpc from the Sun\footnote{Half of the B23 stars lie within 2.4~kpc from the Sun while 95\% of them lie within $\simeq 5.1$~kpc.}, in the range $-3.0\la {\rm [Fe/H]} \la +0.8$. While significantly smaller than many other samples providing photometric metallicities for MW stars \citep[see, e.g.,][]{yang22,xu22,andrae23,martin23}, the B23 sample is carefully selected, homogeneous, and, most importantly for the present application, limited to stars with excellent astrometry from Gaia DR3 \citep[errors on parallax lower than 10\%;][]{dr3_general}.
Moreover, the sample is much larger than those previously used to investigate the metal-poor disc, including the recent thorough analysis of the disc evolution by \citet{aurora22}. 
Here we consider the 685087 stars having valid radial velocity (RV) measures from Gaia DR3 \citep{dr3_rv}, 95\% of these stars having RV uncertainties $\le 5$~km~s$^{-1}$.

In the paper we use the [Fe/H] values obtained by B23 using the \citet{calamida07} [Fe/H]$=f(m_0,(v-y)_0)$ relation, re-calibrated on APOGEE 
DR~17 \citep{apogee_dr17} spectroscopic iron abundances.
For more details on the sample properties we refer the interested reader to B23, where it has been used to derive the metallicity distribution functions (MDF) of several substructures identified by \citet{dodd23}. However, it may be useful to recall here that we adopt the same estimates of the integral of motions and orbital parameters of stars computed by B23 with AGAMA \citet{vasiliev19}, adopting the Galactic potential by \citet{mcmillan17}.

\section{The kinematics of metal-poor disc-like stars}
\label{kinematics}

The metal-poor disc-like component of the local MW population emerges clearly from the B23 data in different ways. Here we display it in a plane of the space parameters that seems especially informative to us, i.e. the one relating the vertical component of the orbital angular momentum $L_Z$ (in km~s$^{-1}$~kpc, with positive values indicating prograde motion, i.e. rotation around the Galactic centre in the same sense as the Sun) to the metallicity [Fe/H]\footnote{An alternative view is provided in Fig.~\ref{fig:LzEn}, below.}, with stars color-coded according the their orbital eccentricity $\epsilon$\footnote{Defined as $\epsilon = (r_{apo}-r_{peri})/(r_{apo}+r_{peri})$, where $r_{apo}$ and $r_{peri}$ are the apogalactic and perigalactic distance, respectively.}.

\begin{figure}[ht!]
\center{
\includegraphics[width=\columnwidth]{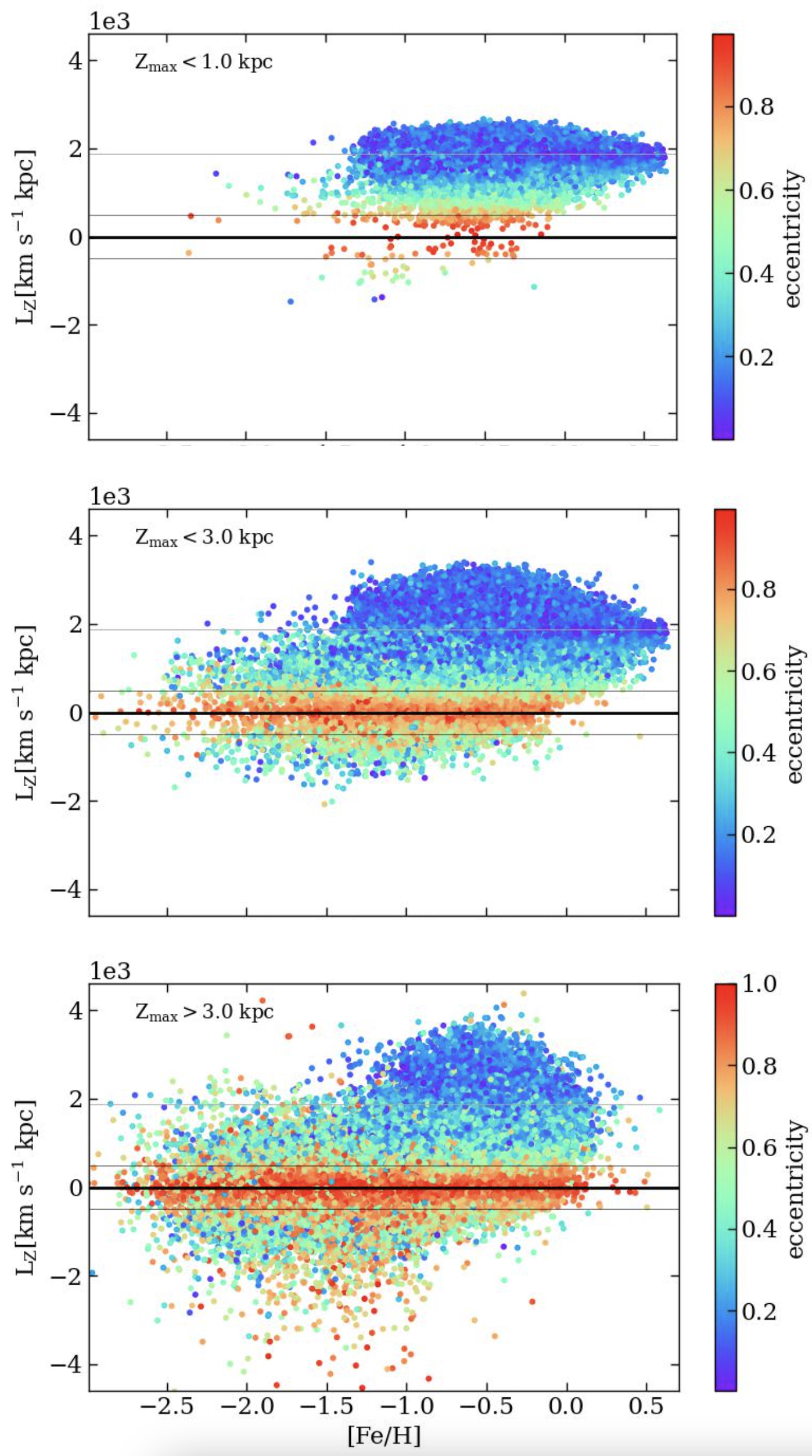}
}
\caption{Metallicity of stars in our sample versus the component of the angular momentum perpendicular to the disc plane for three different slices
in the maximum absolute height over the Galactic plane reached by the star orbit. Points are color-coded according to the orbital eccentricity. The thin black lines are at $L_Z=\pm 500$~km~s$^{-1}$~kpc, while the grey line marks the Z angular momentum of the Sun, $L_Z= 1909$~km~s$^{-1}$~kpc 
in the adopted reference frame. The subsamples plotted in the various panels contain 185178, 580551, and 104530 stars, respecively, from top to bottom.}
\label{fig:LzMet}
\end{figure} 

In Fig.~\ref{fig:LzMet} we show the distibution of stars in this plane for three slices of the sample in $Z_{max}$, that is the maximum absolute height over the Galactic plane reached by a star orbit. This implies that, for example, undisturbed stars with $Z_{max}<1.0$~kpc are permanently confined within this distance from the plane, modulo the uncertainties in the integration of the orbit. The upper panel of Fig.~\ref{fig:LzMet} shows exactly this $Z_{max}<1.0$~kpc subset.

Given the main selection of the B23 sample (in Galactic latitude, $|b|>20.0\degr$), these very-low-$Z_{max}$ stars are confined within a distance range of $\simeq 2.5$~kpc from the Sun.
The B23 sample is dominated by thick disc stars but in this slice the contamination by the more metal-rich and colder thin disc is significant. 
Here stars from both discs are confined in the narrow (i.e., dinamically cold) strip of very low eccentricity and are clustered around $L_Z=L_{Z\sun}$, where $L_{Z\sun}= 1909$~km~s$^{-1}$~kpc in our reference frame. The contribution of thin disc stars (very low eccentricity and low dispersion about $L_{Z\sun}$) is clearly prevalent, especially for [Fe/H]$\ga -0.5$, where the dispersion in $L_Z$ decreases with increasing metallicity. Stars with small net rotation or retrograde are rare\footnote{The three low-eccentricity stars with [Fe/H]$<-1.0$ around $L_Z\simeq -1500$~km~s$^{-1}$~kpc have also virtually the same energy and have properties quite similar to Thamnos \citep{koppelman19} stars. However neither \citet{dodd23} nor B23 associated them to this substructure, due to their extreme $L_Z$ and $\epsilon$ values with respect to the typical ranges adopted to define it.}

In the middle panel the planar slice is extended to $Z_{max}<3.0$~kpc. The bulk of the disc population displays a wider distribution in $L_Z$ due to a larger prevalence of thick disc stars. The stars with $L_Z>L_{Z\sun}$ virtually disappear for [Fe/H]$<-1.4$ (in all the three $Z_{max}$ slices shown in Fig.~\ref{fig:LzMet}), in correspondence to a sharp drop in the MDF of the bulk of thick disc stars \citep[see][]{ivezic08}. The retrograde side of the panel begins to be populated over a wide range of metallicity. A narrow stripe of high-eccentricity stars appears around $L_Z=0.0$~km~s$^{-1}$~kpc. These stars, which are present in different amounts in  all the panels of Fig.~\ref{fig:LzMet}, should be mostly associated with the Gaia-Enceladus/Sausage (GES) remnant \citep{helmi18,belokurov18}, for [Fe/H]$\le-0.5$, while in the most metal-rich regime there is likely a significant contribution from heated disc stars, the so called Splash population \citep{belokurov20}. For [Fe/H]$<-1.4$ a population of moderate eccentricity stars with slower net rotation than the Sun (in both the prograde and retrograde direction) extends to the the metal-poor limit of our sample: these are the metal-poor disc-like stars that will be the subject of the following sections. The prevalence of prograde stars over their retrograde counterparts in this regime is already evident from this panel. The prograde to retrograde ratio for stars with $Z_{max}<3.0$~kpc and [Fe/H]<-1.4 is $P/R=2.58\pm 0.13$.

Finally, in the lower panel ($Z_{max}>3.0$~kpc) the contamination from the thin disc disappears and the bulk of the thick disc population clearly stands out in the range $-1.4\la {\rm [Fe/H]}\la 0.0$, while the high-eccentricity very low rotation component reaches its maximum. In general, stars with [Fe/H]$\la -1.0$ show a significantly more dispersed distribution of $L_Z$ with respect to the $Z_{max}<3.0$~kpc slice, as expected from a halo population. In this metallicity range, the prograde and retrograde sides of the $L_Z$ distribution are roughly symmetric, except for a sparse $L_Z\la -2000$~km~s$^{-1}$~kpc population that includes stars from highly retrograde substructures like, e.g., Sequoia and Thamnos \citep[see][and B23]{myeong19,koppelman19b}. 

Incidentally, we draw the reader's attention to the few tens of [Fe/H]$\le -2.0$ moderate eccentricity stars clustered around the  $L_Z=L_{Z\sun}$ line. These stars are remarkably confined also in terms of total energy, most of them, in particular those with $\epsilon\la 0.5$, being enclosed within $-1.6\times10^5$~km$^2$s$^2\la$~E~$\la-1.3\times10^5$~km$^2$s$^2$, hence they may be part of an unknown substructure\footnote{However, it must be noted that the selection in $Z_{max}$ imposes a lower limit in energy, and $L_Z$ and energy are correlated for strongly rotating, low-eccentricity stars. Hence, the fact that these stars are enclosed in a relatively narrow range in energy may not be particularly significant.}. This group of highly prograde VMP stars seems different from the other known prograde substructures, like e.g. Aleph \citep{naidu20} and Nyx \citep{nyx}, that are much more metal-rich, LMS-1/Wukong \citep{lms-1,naidu20} that has a much lower mean $L_Z$, and Icarus, whose typical stars have orbits confined below $Z_{max}=3.0$ \citep[$\langle Z_{max}\rangle =0.48 \pm 0.59$~kpc,][]{refiorentin21}. Some of the streams identified by \citet{iba19} and \citet{iba21} are similarly prograde and metal-poor \citep[like, e.g., Sv\"ol or Fj\"orm,][]{pristine_streams}, but they are all very narrow and coherent structures, while the group identified here does not show any particular alignment in the sky.

Now, based on the middle panel of Fig.~\ref{fig:LzMet}, we define our sample of disc-like metal-poor stars as those having (a) [Fe/H$\le -1.5$, that is beyond the metal-poor limit of the bulk of the thick disc population, (b) $|L_Z>500|$~km~s$^{-1}$~kpc, that is showing significant prograde or retrograde rotation, following \citet{carollo19}, and (c) $Z_{max}<3.0$~kpc, that is having orbits confined in the proximity of the Galactic plane, as done by \citet{sestito20}. In this way we select 1027 stars, 772 on prograde orbits and 255 on retrograde orbits, confirming the asymmetry already discussed by several authors \citep{sestito20,carter21,carollo23,sotillo23}. The prograde to retrograde ratio is $P/R=3.03\pm 0.22$, assuming Poisson noise, within the range found in simulations of MW-analogues \citep{sestito21,santistevan21,sotillo23}. For example, \citet{santistevan21} finds that the P/R ratio (in mass) in the 12 simulated MW they analyse \citep[from the FIRE suite,][]{hopkins18} ranges from $\simeq 1$ to $\simeq 10$, with a median of $P/R\simeq 2$, nearly constant for [Fe/H]$<-1.5$. 

We further split our sample into a metal-poor (MP, $-2.0<{\rm [Fe/H]}\le -1.5$) and a very metal-poor (VMP, ${\rm [Fe/H]}\le -2.0$) subsamples, containing 834 and 193 stars respectively. The prograde to retrograde ratio is $P/R=3.13\pm 0.25$ in the MP subsample and $P/R=2.64\pm 0.43$ in the VMP one, that is compatible with no difference, within the uncertainties. 

\subsection{Prograde vs. retrograde comparison}
\label{sec:asy}

\begin{figure}[ht!]
\center{
\includegraphics[width=0.9\columnwidth]{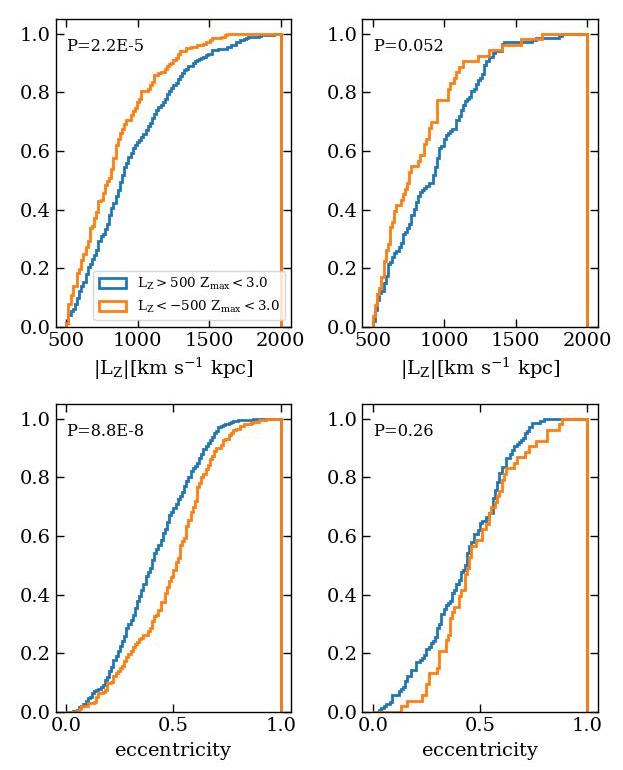}
}
\caption{Comparison of the cumulative distributions of $|L_Z|$ and orbital eccentricity for the  prograde (blue) and retrograde (orange) subsamples of the stars having $|L_Z|>500$~km~s$^{-1}$~kpc and ${\rm Z_{max}}<3.0$~kpc. The comparisons are performed separately for MP (834 stars, left column of panels) and VMP stars (193 stars, right column of panels). In the upper left corner of each panel is reported the probability that the prograde and retrograde samples are drawn from the same parent distribution in the considered quantity, according to a Kolmogorov-Smirnov test.}
\label{fig:cumu}
\end{figure}

Is the asymmetry between the prograde and retrograde metal-poor disc-like populations just a matter of numbers of stars or does it involve other properties? To investigate this issue, in Fig.~\ref{fig:cumu} we compare the distribution of $|L_Z|$ and orbital eccentricity of the prograde and retrograde stars on disc-like orbits, as defined above, separately for the MP (left panels) and VMP (right panels) subsamples.

The prograde MP population is remarkably more skewed toward high $|L_Z|$ (strong planar rotation) and low eccentricity than its retrograde counterpart. The statistical significance of the detected differences is very high in both cases, according to the Kolmogorov-Smirnov test. The same differences are observed for the VMP sample but in this case the statistical significance is only marginal for the comparison of the $|L_Z|$ distributions and negligible for the eccentricity distributions: this is likely owed to the much smaller sample size with respect to the MP case. However, the fact that the $|L_Z|$ value at which the maximum difference between the prograde and retrograde distributions is basically the same for the MP ($|L_Z|$=878~km~s$^{-1}$~kpc) and VMP ($|L_Z|$=912~km~s$^{-1}$~kpc) cases strongly support the hypothesis that the two subsamples have the same behaviour in this respect. Concerning eccentricity, the largest difference in the distributions is found at $\epsilon=0.16$ for the VMP and at $\epsilon=0.25$ for the MP, in both cases an excess of low eccentricity orbits among the prograde stars w.r.t. their retrograde counterparts.

\begin{figure}[ht!]
\center{
\includegraphics[width=0.9\columnwidth]{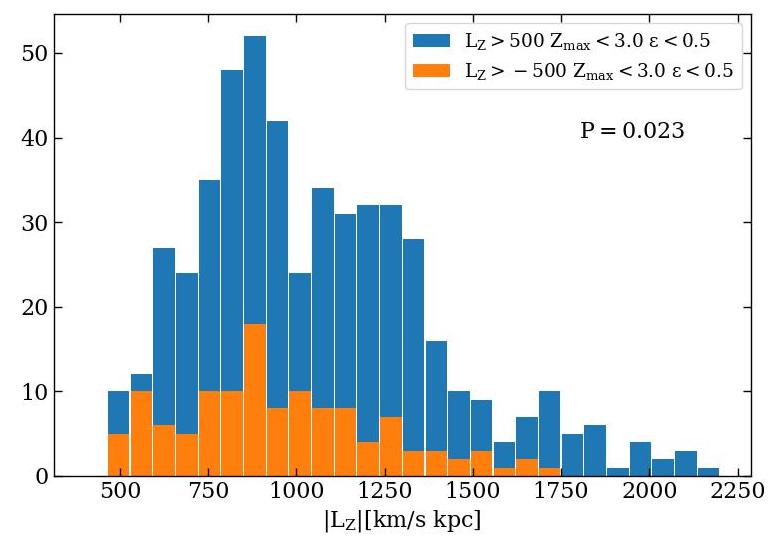}
}
\caption{Comparison of the $|L_Z|$ distributions for the samples of prograde (blue) and retrograde (orange) stars with disc-like orbits (as defined in Fig.~\ref{fig:cumu}) and low orbital eccentricity ($\epsilon<0.5$). All stars with [Fe/H]$\le -1.5$ are included in the comparison.
The probability that the two samples are drawn from the same parent population in $|L_Z|$ according to a Kolmogorov-Smirnov test is also reported.}
\label{fig:Lz_lowecc}
\end{figure} 

To better understand the reasons of the differences between the prograde and retrograde metal-poor disc populations, in Fig.~\ref{fig:Lz_lowecc} we compare again the $|L_Z|$ distribution but this time in differential form and merging together the MP and VMP samples and restricting to stars on moderate eccentricity orbits ($\epsilon<0.5$; 517 prograde and 124 retrograde stars). It is interesting to note that the two distributions have similar characteristic shapes, with a main peak around $|L_Z|$=900~km~s$^{-1}$~kpc and a long declining tail toward higher $|L_Z|$ values, but the distribution of the prograde stars displays an additional broad component peaking at  $|L_Z|\sim 1200$~km~s$^{-1}$~kpc. This is similar 
to the metal weak thick disc component discussed by \citet{carollo19} but in a different metallicity regime, as the stars they attribute to that component have $-1.2<{\rm[Fe/H]}<-0.6$ while those considered here have [Fe/H]$\le -1.5$. From the perspective of the metallicity range our sample is more akin to the metal weak thick disc as defined and studied by \citet{li_zhao17}.

\subsection{The early evolution of the MW disc}
\label{sec:aurora}

In a recent paper \citet[][hereafter BK22]{aurora22}, based on evidence drawn from the analysis of a sample selected from the APOGEE~DR17 dataset \citep{apogee_dr17}, presented an intriguing view of the evolution of the Galactic disc from an initial chaotic relatively metal-poor phase (that they dub "Aurora") to the present-day cold disc stage through a rapid spinning-up phase occurred while the mean metallicity of the stars grew from [Fe/H]$\simeq -1.5$ to [Fe/H]$\simeq -0.9$. They provided both observational and theoretical support for a phase transition between a regime in which gas accretion in the pristine Galaxy occurs mainly in an irregular and turbulent fashion from narrow filamentary flows to a regime of more ordered accretion via cooling flows from a hot gaseous halo, leading to the formation of a coherently rotating disc (see also \citealt{rix22}, \citealt{xiang_rix22} and \citealt{conroy22}; for a study of the process from the theoretical side see \citealt{dillamore23}, and references therein). 
Recently \citet{chandra23} presented a very similar view of the evolution of the discy components of the MW from the analysis of a large set of giant stars with photometric metallicities and [$\alpha$/Fe] ratios and of simulations of MW-analogues from the IllustrisTNG suite \citep{pillepich18}.
In this section we discuss how the results presented in this paper fit into the BK22 framework. Here it is worth noting that in a galaxy formation process driven by hierarchical merging the definition of "in-situ" becomes blurred at early times, before a relatively stable gaseous disc is settled.
Hence, in the MP and VMP regimes on which we are mainly focusing our attention, in-situ broadly denotes the early epoch in which the first building blocks were merging together to form the seed of our own Galaxy \citep[see, e.g, BK22,][]{sestito21,chandra23}.

\begin{figure}[ht!]
\center{
\includegraphics[width=\columnwidth]{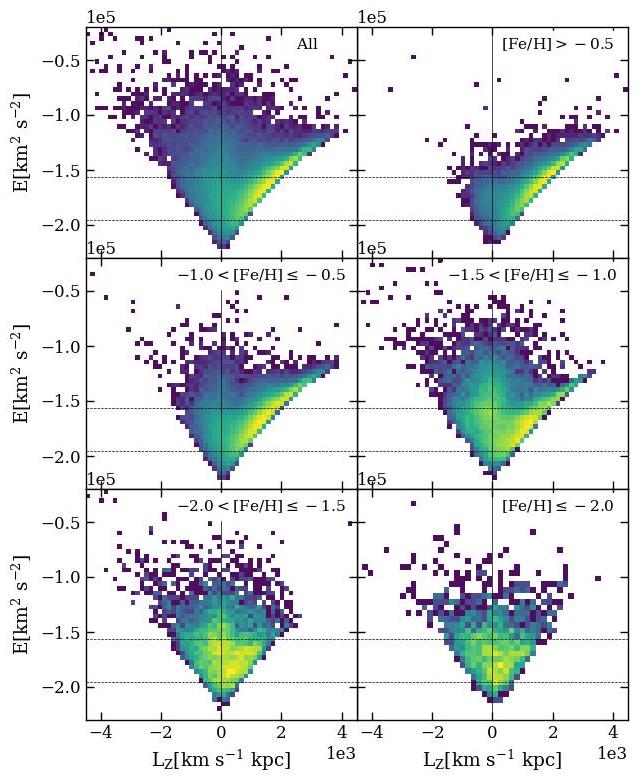}
}
\caption{2D histograms of stars in our sample in the angular momentum versus orbital energy plane. The distribution is shown for the entire sample  (upper left panel) and for different metallicity slices, from the most metal-rich (upper right panel) to the most metal-poor (lower right panel). The vertical continuous line separate prograde from retrograde orbits (to the right and to the left of the line, respectively). The two dotted horizontal lines enclose the energy slice that should be dominated by in-situ stars, following BK22.}
\label{fig:LzEn}
\end{figure} 

Being interested to stars born in-situ in the Galaxy, BK22 effectively remove the stars that have been most likely accreted during merger events using the remarkable discriminating power of the [Al/Fe] abundance \citep{hawkins15}. Using [Al/Fe] and energy constraints they finally select a local sample of $\simeq 80000$ in-situ stars, most of which lying in the range $-1.7\la {\rm [Fe/H]}\la 0.5$. 
It turns out that the vast majority of in-situ stars selected by BK22 are confined in the range of orbital energy $-0.75\times 10^5$~km$^2$~s$^{-2}<E<-0.4\times 10^5$~km$^2$~s$^{-2}$, hence we can use a cut in energy to draw a sub-sample that should be dominated by in-situ stars from our sample, roughly approximating the BK22 selection. This will allow us to re-consider the issue discussed in Sect.~\ref{sec:asy} from a different perspective, checking if we obtain consistent results and how they fit into the scenario depicted by BK22.

Since the Galactic potentials adopted by us and by BK22 are different, the energy scale is also different. We used 51460 stars in common between our sample and the dataset from which they draw the orbital parameters and integrals of motion for their stars \citep[the astroNN\footnote{\url{https://www.sdss.org/dr18/data_access/value-added-catalogs/?vac_id=85}} value added catalogue,][]{astronn} to fit a linear relation translating energy values from one scale to the other, $E_{t.w.}=1.097E_{BK22} -1.12\times 10^5$. The relation is linear (Pearson's linear correlation coefficient $\rho=0.972$) and very tight (r.m.s =$0.03\times 10^5$~km$^2$~s$^{-2}$), especially in the range of energies $-2.3\times 10^5~{\rm km^2~s^{-2}}\la E\la -1.2\times 10^5~{\rm km^2~s^{-2}}$, that widely encloses the most relevant range for the following analysis.
With this relation the energy window for in-situ stars in our scale is $-1.95\times 10^5$~km$^2$~s$^{-2}<E<-1.56\times 10^5$~km$^2$~s$^{-2}$. This window is represented by two parallel dotted lines in each panel of Fig.~\ref{fig:LzEn}, where the wealth of dynamical structures in our sample is displayed in the $L_Z$ - E plane sliced by metallicity. In the lowest panels the prograde-retrograde asymmetry in fast-rotating MP and VMP stars can be clearly appreciated. This figure allows a clear and direct comparison with BK22, see their Figure~1, in particular.

\begin{figure}[ht!]
\center{
\includegraphics[width=\columnwidth]{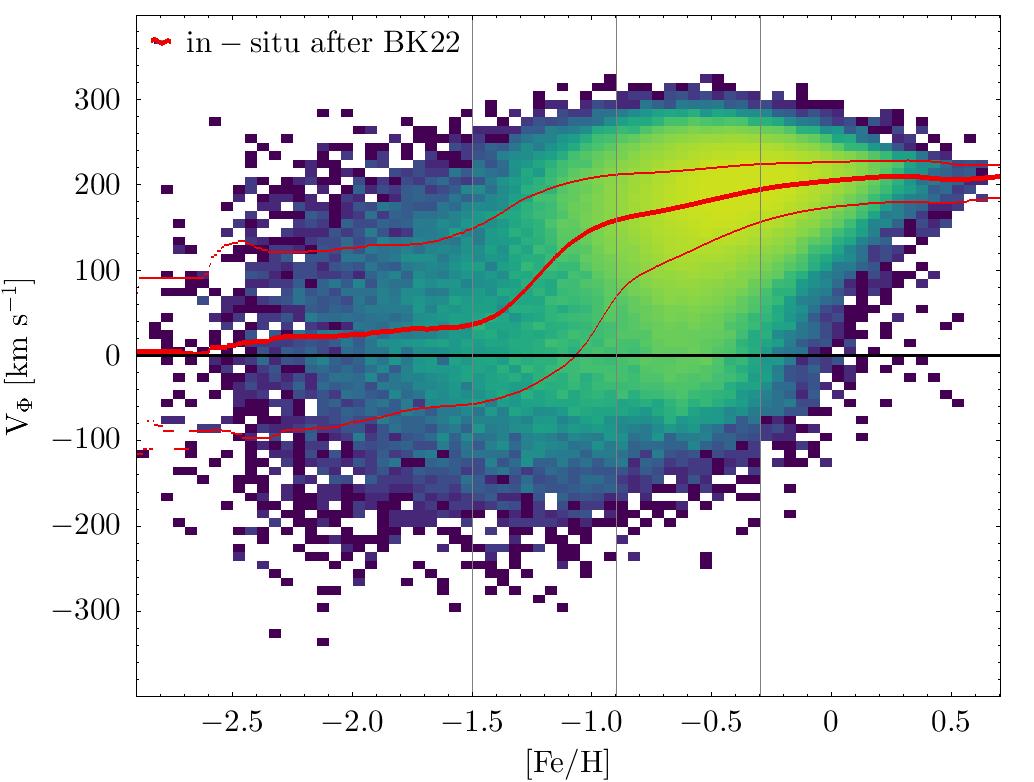}
}
\caption{Distribution in the metallicity versus azimuthal velocity for the 446754 stars in our sample that should have been born in-situ according the the selection in orbital energy shown by BK22 in their Figure~1. The thick red line traces the median $V_{\Phi}$ of the distribution, while the thinner lines traces the 5th and 95th percentile. The vertical grey lines mark the transition points suggested by BK22, [Fe/H]$=-0.3$ and  [Fe/H]$=-0.9$, as well as the lower limit of the metallicity range in which the same trend is discussed in BK22, [Fe/H]$=-1.5$ (see their Fig.~4).}
\label{fig:FeVphi}
\end{figure} 

The [Fe/H] vs. $V_{\Phi}$ distribution of the 446754 stars in our sample enclosed in this in-situ energy window is displayed in Fig.~\ref{fig:FeVphi}, that is fully analogue to the lower right panel of Fig.~3 of BK22. Here we nicely reproduce the BK22 result with a much larger sample\footnote{It is worth noting here that the adopted energy cuts were effective in removing most of the stars associated to GES and other high-energy substructures (see B23) and allowed us to obtain a cleaner reproduction of the BK22 result, improving the consistency between the two analyses. However the general trends of the median and of the dispersion shown in Fig.~\ref{fig:FeVphi}  emerge quite clearly also when the entire B23 sample is considered.} \citep[see also][]{chandra23}. At the earliest epochs ([Fe/H]$\le -1.5$) the net rotation of in-situ stars is very low, or null, and the dispersion is high ($\simeq 150$~km~s$^{-1}$). In the range between [Fe/H]$\ga -1.5$ and [Fe/H]$\la -0.9$ there is a strong increase in the median azimuthal velocity, up to about $V_{\Phi}=130$~km~s$^{-1}$. Then the increase in $V_{\Phi}$ becomes much milder, flattening out at $V_{\Phi}\simeq 210$~km~s$^{-1}$ for [Fe/H]$\ga 0.0$, while the dispersion shrinks to a few tens of km~s$^{-1}$. The only small difference with respect to BK22 is that the median azimuthal velocity in our sample of putative in-situ stars seems to reach  $V_{\Phi}=0.0$~km~s$^{-1}$ only for [Fe/H]$\la -2.5$, while it is $V_{\Phi}\simeq 20$~km~s$^{-1}$ at [Fe/H]$\simeq -2.0$ and $V_{\Phi}\simeq 40$~km~s$^{-1}$ at [Fe/H]$\simeq -1.5$ \citep[in excellent agreement with the median trend measured in MW-like galaxy models by][see, in particular, their Fig.~1]{dillamore23}\footnote{In a previous contribution, \citet{dillamore_bar} suggested that bar resonances may drive stars from the inner Galaxy to the Solar neighbourhood on nearly circular orbits. However, in their recent analysis, \citet{yuan23} concluded that this mechanism is not able to produce the observed local population of metal-poor stars on prograde disc-like orbits.}, possibly suggesting a mild spinning-up also at very early epochs or, in other words, a pristine prevalence of stars rotating in the sense that will become largely dominant in the mature phases of the disc(s) evolution.  

\begin{figure}[ht!]
\center{\includegraphics[width=\columnwidth]{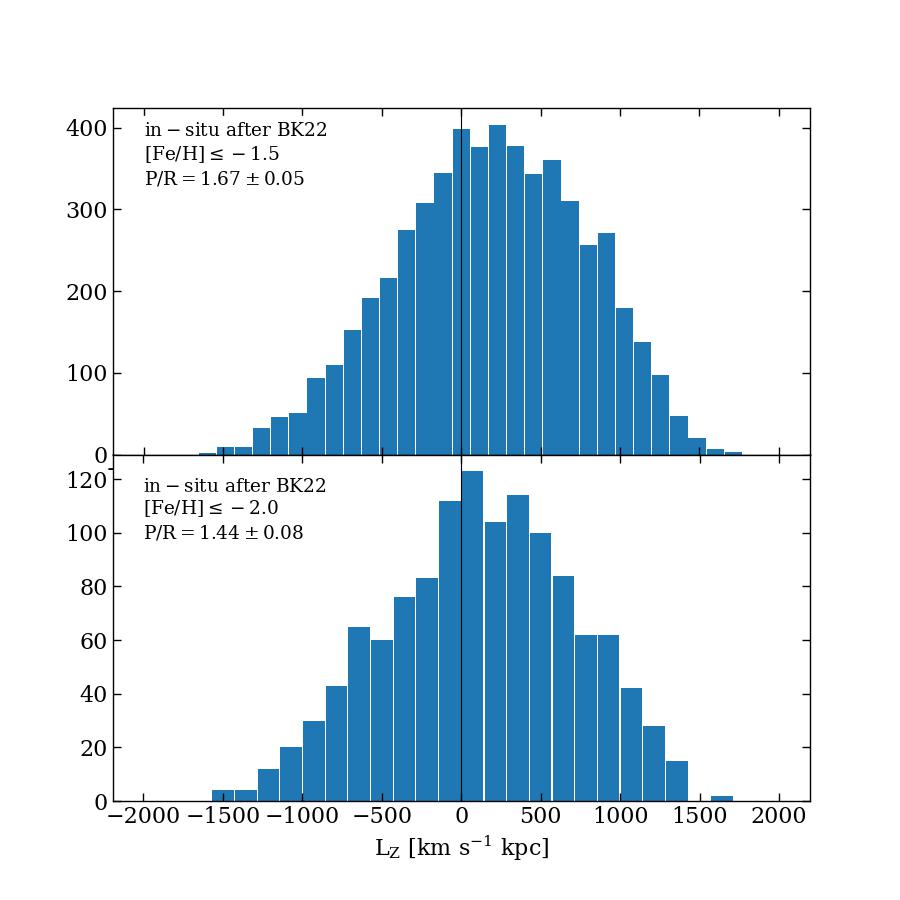}}
\caption{$L_Z$ distributions of the in-situ stars with [Fe/H]$\le -1.5$
(MP+VMP; upper panel) and with [Fe/H]$\le -2.0$ (VMP; lower panel).
The prograde to retrograde ratio$P/R$ in reported in each panel for the considered subsample.
}
\label{fig:VphiMP}
\end{figure} 

This conclusion is supported by the distributions in $L_Z$
of the in-situ stars with [Fe/H]$\le -1.5$  and [Fe/H]$\le -2.0$ shown in Fig.~\ref{fig:VphiMP}. The prograde/retrograde asymmetry is already in place in these metallicity ranges also for in-situ stars, either by number, with 3394 prograde stars and 2033 retrograde stars for the MP+VMP sample and  734 prograde stars and 509 retrograde stars for the VMP sample,
and by shape. In particular, in the MP+VMP case the retrograde side of the distribution is smooth and (half-)bell-shaped while the prograde side has much more power at  high $|L_Z|$ values (the same is true if one looks at the high $V_{\Phi}$ distribution).
The difference between the two halves of the distribution is significant in both metallicity ranges. According to a Kolmogorov-Smirnov test the probability that the positive and negative side of the  distributions shown in Fig.~\ref{fig:VphiMP} are drawn from the same parent distribution of $|L_Z|$ is $P=6.6\times 10^{-29}$ for the MP+VMP sample and $P=0.010$ for the smaller VMP sample.

\section{Summary and Conclusions}
\label{conclu}

We have used the dataset by B23 to study the properties of the MP (-2.0<[Fe/H]$\le -1.5$) and VMP ([Fe/H]$\le -2.0$) stars having disc-like orbits in the vicinity of the Sun. We have confirmed the significant prevalence of prograde over retrograde disc-like stars over the entire metallicity range considered. We have also shown that in both MP and VMP subsamples,  prograde stars with high planar rotation and low eccentricity are much more frequent than retrograde stars with the same properties. The differences in the $|L_Z|$ and $\epsilon$ distributions of the MP sample have very high statistical significance. 
For the first time, a qualitative asymmetry seems to emerge between the prograde and retrograde metal-poor disc-like populations, in addition to the quantitative one already noted and discussed by other authors \citep[see][and references therein]{sestito20,carter21,carollo23}. This have been made possible by the larger sample of MP stars with excellent measures of the 6-D phase space parameters that we were able to analyse here, with respect to previous studies.

Merging together the MP and VMP stars and restricting to moderate eccentricity stars ($\epsilon<0.5$) we found that the $|L_Z|$ distribution of prograde stars appears bimodal, with a secondary peak at relatively high planar rotation that seems to lack a counterpart in the retrograde sample, possibly hinting at the presence of two different disc-like components on the prograde side.   

A sizeable fraction of metal-poor stars on disc-like orbits have been found by various authors in recent physically-motivated numerical simulations of galaxies similar to the MW  \citep{sestito21,santistevan21,carollo23,sotillo23}, with non-negligible galaxy-to-galaxy differences. For instance, \citet{sotillo23} found that the fraction of VMP stars in the cold disc of the model galaxies they analysed may vary from about 5-10\% up to about 40\% in the most extreme cases, with typical values around 20\%. All these studies agree that accretion is the dominant channel for the build up of the extremely metal-poor disc-like population ([Fe/H]$<-2.5$) but \citet{carollo23} highlighted that in situ formation may also contribute and  \citet{sotillo23} found that the fraction of disc stars formed in-situ decreases with decreasing metallicity
\citep[in agreement with][]{conroy19}, from about 70\% around [Fe/H]$\simeq -1.0$ to less than 50\% for [Fe/H]$\la -3.0$, for their "warm disc" component, the one more akin to the sample studied here. In a recent analysis of 24 Milky Way analogues from the AURIGA simulation \citep{grand17} aimed at studying the origin of the thick and thin discs in an extragalactic context, \citet{pinna23} concluded that thick discs mostly formed in-situ and at very early times, albeit with a significant contribution of satellite mergers.
On the other hand, \citet{santistevan21} found that the typical most important contributor to the prevalence of prograde over retrograde stars in their set of MW-analogues is a single early galaxy merger, who may play also a significant role in the build up of the MW disc \citep[see also][]{chandra23}. Similarly to \citet{santistevan21}, \citet{sestito21}, who studied the origin of VMP stars in five high resolution of MW-like galaxies from the NIHAO-UHD simulations \citep{buck20}, concluded that VMP disc-like stars are mainly contributed by major building blocks during the early assembly phase of the MW, but can be possibly brought in also during later merging events, when prograde satellites are more easily dragged into the already-formed MW disc than retrograde ones.

In any case, in the range of metallicity explored in the present study, a non-negligible very early component of the metal-poor disc is expected, born either in the initial most massive seed of the MW assembly (in situ) or in one of the building blocks that merged with that seed at very early epochs. We speculate that the population producing the prograde/retrograde differences in the $|L_Z|$ and $\epsilon$ distributions shown in Fig.~\ref{fig:cumu} as well as the high $|L_Z|$ peak that we identified in the $|L_Z|$ distribution of prograde stars with $\epsilon <0.5$ shown in Fig.~\ref{fig:Lz_lowecc} can be such a pristine component. 

This hypothesis is supported by the results we obtained from the analysis of a large sub-sample that should be dominated by stars classified as
in-situ, following the approach of BK22.
We were able to reproduce very well and with a  much larger sample (446754 stars vs. the $\simeq 79500$ used by BK22), the trend of the median and of the dispersion in azimuthal velocity with metallicity first highlighted by BK22 and interpreted by these authors as the transition from a pristine very chaotic phase
(Aurora, [Fe/H]$\la -1.5$) to a spinning-up and more ordered gas accretion phase that lead to the formation of the present-day disc \citep[see also][]{rix22,conroy22,dillamore23,chandra23,semenov23a}. The larger sample also allowed us to trace the trend of median azimuthal velocity with [Fe/H] to much lower metallicity than BK22 (beyond [Fe/H]=-2.0, while they reach [Fe/H]$\simeq $-1.5) and also than \citet{chandra23}, who limit their analysis to [Fe/H]=-2.0.
We showed that, even in the Aurora phase, there was a clear prevalence of 
prograde stars over retrograde ones and the $|L_Z|$ distribution of prograde stars were significantly more skewed toward high values (i.e., stronger rotation) than its retrograde counterpart. 

In this context, it seems worth mentioning the results of two very recent analyses of large samples of local stars based on two different sets of photometric metallicities.
\citep{zhang23} suggested that metal-poor stars with disc-like orbits are not a special MW component in the metal-poor regime but just the subset of halo stars that is expected to have prograde and low $Z_{max}$ orbits, with the prevalence of prograde stars being due to a slightly prograde halo component. They show that the distribution of $\epsilon$ of $Z_{max}\le 3.0$~kpc in their sample begins to show a prevalence of low eccentricity orbits only for [Fe/H]$>-1.3$, while at lower metallicity it appears indistinguishable from that of a simulated non-rotating halo population. Here we note that, in our MP sample with $Z_{max}\le 3.0$~kpc  the prograde population shows a clear low-eccentricity component (peaking at $\epsilon\simeq 0.3-0.4$) that has no counterpart in the retrograde population. Hence, the results shown in Fig.~\ref{fig:cumu} remain valid even if the constraint on |$L_Z$| adopted there is dropped, an occurrence that seems difficult to explain just with a non rotating halo.
In line with this, \citep{hong23} traces a disc-like component in the local Galaxy down to [Fe/H]$\la -3.0$, suggesting the presence of a forming disc at very early epochs.

In summary, our results suggest that, in the metal-poor regime, in addition to a symmetric accreted disc-like component, the seed of what lately evolved into the Galactic 
disc was already present at very early epochs, and is clearly visible in the surroundings of the Sun, at least when a large sample of stars with accurate metallicity, distance and kinematic measures is considered. As a final consideration, we feel that since last decade we are just beginning to trace the early assembly phase of the MW disc. As usual, scientists trying to tackle the problem from different perspectives provide different valuable pieces of information that take some time to find their place together into a global picture. We think that the present contribution may be a useful step in bridging two approaches to the early formation of the disc, one focusing on relatively small samples of very and extremely metal-poor stars, like, e.g., \citet{sestito19,sestito20,cordoni21,carollo23}, with the possiblity of deeper insight from detailed chemical composition \citep{dovgal23}, and the other trying to reconstruct the main phases of the evolutionary path with large samples limited to a metallicity range ([Fe/H]$\ga -1.5$) where local disc stars are more abundant \citep[like, e.g., BK22,][]{xiang_rix22,chandra23}.

\begin{acknowledgements}

MB, PM, AB and DM acknowledge the support to activities related to the ESA/\Gaia mission by the Italian Space Agency (ASI) through contract 2018-24-HH.0 and its addendum 2018-24-HH.1-2022 to the National Institute for Astrophysics (INAF). MB, EC, DM and AM acknowledge the support to this study by the PRIN INAF 2023 grant ObFu {\em CHAM - Chemo-dynamics of the Accreted Halo of the Milky Way} (P.I. M. Bellazzini) and by the project {\em LEGO – Reconstructing the building blocks of the Galaxy by chemical tagging} (P.I. A. Mucciarelli), granted by the Italian MUR through contract PRIN 2022LLP8TK\_001.

MB is grateful to R. Pascale for his help in the production of figures with Python. We are grateful to an anonymous Referee for constructive comments that helped us to improve the paper.

This work has made use of data from the European Space Agency (ESA) mission \Gaia (https://www.cosmos.esa.int/Gaia), processed by the \Gaia Data Processing and Analysis Consortium (DPAC, https://www.cosmos.esa.int/web/Gaia/dpac/consortium). Funding for the DPAC has been provided by national institutions, in particular the institutions participating in the \Gaia Multilateral Agreement.

In this analysis we made use of TOPCAT (http://www.starlink.ac.uk/topcat/, \citealt{Taylor2005}).

\end{acknowledgements}


\bibliographystyle{aa} 
\bibliography{refs} 

\begin{thebibliography}{62}
\expandafter\ifx\csname natexlab\endcsname\relax\def\natexlab#1{#1}\fi

\bibitem[{{Abdurro'uf} {et~al.}(2022){Abdurro'uf}, {Accetta}, {Aerts}, {Silva
  Aguirre}, {Ahumada}, {Ajgaonkar}, {Filiz Ak}, {Alam}, {Allende Prieto},
  {Almeida}, {Anders}, {Anderson}, {Andrews}, {Anguiano}, {Aquino-Ort{\'\i}z},
  {Arag{\'o}n-Salamanca}, {Argudo-Fern{\'a}ndez}, {Ata}, {Aubert},
  {Avila-Reese}, {Badenes}, {Barb{\'a}}, {Barger}, {Barrera-Ballesteros},
  {Beaton}, {Beers}, {Belfiore}, {Bender}, {Bernardi}, {Bershady}, {Beutler},
  {Bidin}, {Bird}, {Bizyaev}, {Blanc}, {Blanton}, {Boardman}, {Bolton},
  {Boquien}, {Borissova}, {Bovy}, {Brandt}, {Brown}, {Brownstein}, {Brusa},
  {Buchner}, {Bundy}, {Burchett}, {Bureau}, {Burgasser}, {Cabang}, {Campbell},
  {Cappellari}, {Carlberg}, {Wanderley}, {Carrera}, {Cash}, {Chen}, {Chen},
  {Cherinka}, {Chiappini}, {Choi}, {Chojnowski}, {Chung}, {Clerc}, {Cohen},
  {Comerford}, {Comparat}, {da Costa}, {Covey}, {Crane}, {Cruz-Gonzalez},
  {Culhane}, {Cunha}, {Dai}, {Damke}, {Darling}, {Davidson}, {Davies},
  {Dawson}, {De Lee}, {Diamond-Stanic}, {Cano-D{\'\i}az}, {S{\'a}nchez},
  {Donor}, {Duckworth}, {Dwelly}, {Eisenstein}, {Elsworth}, {Emsellem},
  {Eracleous}, {Escoffier}, {Fan}, {Farr}, {Feng}, {Fern{\'a}ndez-Trincado},
  {Feuillet}, {Filipp}, {Fillingham}, {Frinchaboy}, {Fromenteau}, {Galbany},
  {Garc{\'\i}a}, {Garc{\'\i}a-Hern{\'a}ndez}, {Ge}, {Geisler}, {Gelfand},
  {G{\'e}ron}, {Gibson}, {Goddy}, {Godoy-Rivera}, {Grabowski}, {Green},
  {Greener}, {Grier}, {Griffith}, {Guo}, {Guy}, {Hadjara}, {Harding},
  {Hasselquist}, {Hayes}, {Hearty}, {Hern{\'a}ndez}, {Hill}, {Hogg},
  {Holtzman}, {Horta}, {Hsieh}, {Hsu}, {Hsu}, {Huber}, {Huertas-Company},
  {Hutchinson}, {Hwang}, {Ibarra-Medel}, {Chitham}, {Ilha}, {Imig}, {Jaekle},
  {Jayasinghe}, {Ji}, {Johnson}, {Jones}, {J{\"o}nsson}, {Katkov}, {Khalatyan},
  {Kinemuchi}, {Kisku}, {Knapen}, {Kneib}, {Kollmeier}, {Kong}, {Kounkel},
  {Kreckel}, {Krishnarao}, {Lacerna}, {Lane}, {Langgin}, {Lavender}, {Law},
  {Lazarz}, {Leung}, {Leung}, {Lewis}, {Li}, {Li}, {Lian}, {Liang}, {Lin},
  {Lin}, {Lin}, {Lintott}, {Long}, {Longa-Pe{\~n}a}, {L{\'o}pez-Cob{\'a}},
  {Lu}, {Lundgren}, {Luo}, {Mackereth}, {de la Macorra}, {Mahadevan},
  {Majewski}, {Manchado}, {Mandeville}, {Maraston}, {Margalef-Bentabol},
  {Masseron}, {Masters}, {Mathur}, {McDermid}, {Mckay}, {Merloni},
  {Merrifield}, {Meszaros}, {Miglio}, {Di Mille}, {Minniti}, {Minsley},
  {Monachesi}, {Moon}, {Mosser}, {Mulchaey}, {Muna}, {Mu{\~n}oz}, {Myers},
  {Myers}, {Nadathur}, {Nair}, {Nandra}, {Neumann}, {Newman}, {Nidever},
  {Nikakhtar}, {Nitschelm}, {O'Connell}, {Garma-Oehmichen}, {Luan Souza de
  Oliveira}, {Olney}, {Oravetz}, {Ortigoza-Urdaneta}, {Osorio}, {Otter},
  {Pace}, {Padilla}, {Pan}, {Pan}, {Parikh}, {Parker}, {Peirani}, {Pe{\~n}a
  Ram{\'\i}rez}, {Penny}, {Percival}, {Perez-Fournon}, {Pinsonneault},
  {Poidevin}, {Poovelil}, {Price-Whelan}, {B{\'a}rbara de Andrade Queiroz},
  {Raddick}, {Ray}, {Rembold}, {Riddle}, {Riffel}, {Riffel}, {Rix}, {Robin},
  {Rodr{\'\i}guez-Puebla}, {Roman-Lopes}, {Rom{\'a}n-Z{\'u}{\~n}iga}, {Rose},
  {Ross}, {Rossi}, {Rubin}, {Salvato}, {S{\'a}nchez}, {S{\'a}nchez-Gallego},
  {Sanderson}, {Santana Rojas}, {Sarceno}, {Sarmiento}, {Sayres}, {Sazonova},
  {Schaefer}, {Schiavon}, {Schlegel}, {Schneider}, {Schultheis}, {Schwope},
  {Serenelli}, {Serna}, {Shao}, {Shapiro}, {Sharma}, {Shen}, {Shetrone}, {Shu},
  {Simon}, {Skrutskie}, {Smethurst}, {Smith}, {Sobeck}, {Spoo}, {Sprague},
  {Stark}, {Stassun}, {Steinmetz}, {Stello}, {Stone-Martinez},
  {Storchi-Bergmann}, {Stringfellow}, {Stutz}, {Su}, {Taghizadeh-Popp},
  {Talbot}, {Tayar}, {Telles}, {Teske}, {Thakar}, {Theissen}, {Tkachenko},
  {Thomas}, {Tojeiro}, {Hernandez Toledo}, {Troup}, {Trump}, {Trussler},
  {Turner}, {Tuttle}, {Unda-Sanzana}, {V{\'a}zquez-Mata}, {Valentini},
  {Valenzuela}, {Vargas-Gonz{\'a}lez}, {Vargas-Maga{\~n}a}, {Alfaro},
  {Villanova}, {Vincenzo}, {Wake}, {Warfield}, {Washington}, {Weaver},
  {Weijmans}, {Weinberg}, {Weiss}, {Westfall}, {Wild}, {Wilde}, {Wilson},
  {Wilson}, {Wilson}, {Wolf}, {Wood-Vasey}, {Yan}, {Zamora}, {Zasowski},
  {Zhang}, {Zhao}, {Zheng}, {Zheng}, \& {Zhu}}]{apogee_dr17}
{Abdurro'uf}, {Accetta}, K., {Aerts}, C., {et~al.} 2022, \apjs, 259, 35

\bibitem[{{Andrae} {et~al.}(2023){Andrae}, {Rix}, \& {Chandra}}]{andrae23}
{Andrae}, R., {Rix}, H.-W., \& {Chandra}, V. 2023, \apjs, 267, 8

\bibitem[{{Beers} {et~al.}(2014){Beers}, {Norris}, {Placco}, {Lee}, {Rossi},
  {Carollo}, \& {Masseron}}]{beers2014}
{Beers}, T.~C., {Norris}, J.~E., {Placco}, V.~M., {et~al.} 2014, \apj, 794, 58

\bibitem[{{Bellazzini} {et~al.}(2023){Bellazzini}, {Massari}, {De Angeli},
  {Mucciarelli}, {Bragaglia}, {Riello}, \& {Montegriffo}}]{B23}
{Bellazzini}, M., {Massari}, D., {De Angeli}, F., {et~al.} 2023, \aap, 674,
  A194

\bibitem[{{Belokurov} {et~al.}(2018){Belokurov}, {Erkal}, {Evans}, {Koposov},
  \& {Deason}}]{belokurov18}
{Belokurov}, V., {Erkal}, D., {Evans}, N.~W., {Koposov}, S.~E., \& {Deason},
  A.~J. 2018, \mnras, 478, 611

\bibitem[{{Belokurov} \& {Kravtsov}(2022)}]{aurora22}
{Belokurov}, V. \& {Kravtsov}, A. 2022, \mnras, 514, 689

\bibitem[{{Belokurov} {et~al.}(2020){Belokurov}, {Sanders}, {Fattahi}, {Smith},
  {Deason}, {Evans}, \& {Grand}}]{belokurov20}
{Belokurov}, V., {Sanders}, J.~L., {Fattahi}, A., {et~al.} 2020, \mnras, 494,
  3880

\bibitem[{{Buck} {et~al.}(2020){Buck}, {Obreja}, {Macci{\`o}}, {Minchev},
  {Dutton}, \& {Ostriker}}]{buck20}
{Buck}, T., {Obreja}, A., {Macci{\`o}}, A.~V., {et~al.} 2020, \mnras, 491, 3461

\bibitem[{{Calamida} {et~al.}(2007){Calamida}, {Bono}, {Stetson}, {Freyhammer},
  {Cassisi}, {Grundahl}, {Pietrinferni}, {Hilker}, {Primas}, {Richtler},
  {Romaniello}, {Buonanno}, {Caputo}, {Castellani}, {Corsi}, {Ferraro},
  {Iannicola}, \& {Pulone}}]{calamida07}
{Calamida}, A., {Bono}, G., {Stetson}, P.~B., {et~al.} 2007, \apj, 670, 400

\bibitem[{{Carollo} {et~al.}(2019){Carollo}, {Chiba}, {Ishigaki}, {Freeman},
  {Beers}, {Lee}, {Tissera}, {Battistini}, \& {Primas}}]{carollo19}
{Carollo}, D., {Chiba}, M., {Ishigaki}, M., {et~al.} 2019, \apj, 887, 22

\bibitem[{{Carollo} {et~al.}(2023){Carollo}, {Christlieb}, {Tissera}, \&
  {Sillero}}]{carollo23}
{Carollo}, D., {Christlieb}, N., {Tissera}, P.~B., \& {Sillero}, E. 2023, \apj,
  946, 99

\bibitem[{{Carter} {et~al.}(2021){Carter}, {Conroy}, {Zaritsky}, {Ting},
  {Bonaca}, {Naidu}, {Johnson}, {Cargile}, {Caldwell}, {Speagle}, \&
  {Han}}]{carter21}
{Carter}, C., {Conroy}, C., {Zaritsky}, D., {et~al.} 2021, \apj, 908, 208

\bibitem[{{Chandra} {et~al.}(2023){Chandra}, {Semenov}, {Rix}, {Conroy},
  {Bonaca}, {Naidu}, {Andrae}, {Li}, \& {Hernquist}}]{chandra23}
{Chandra}, V., {Semenov}, V.~A., {Rix}, H.-W., {et~al.} 2023, arXiv e-prints,
  arXiv:2310.13050

\bibitem[{{Conroy} {et~al.}(2019){Conroy}, {Naidu}, {Zaritsky}, {Bonaca},
  {Cargile}, {Johnson}, \& {Caldwell}}]{conroy19}
{Conroy}, C., {Naidu}, R.~P., {Zaritsky}, D., {et~al.} 2019, \apj, 887, 237

\bibitem[{{Conroy} {et~al.}(2022){Conroy}, {Weinberg}, {Naidu}, {Buck},
  {Johnson}, {Cargile}, {Bonaca}, {Caldwell}, {Chandra}, {Han}, {Johnson},
  {Speagle}, {Ting}, {Woody}, \& {Zaritsky}}]{conroy22}
{Conroy}, C., {Weinberg}, D.~H., {Naidu}, R.~P., {et~al.} 2022, arXiv e-prints,
  arXiv:2204.02989

\bibitem[{{Cordoni} {et~al.}(2021){Cordoni}, {Da Costa}, {Yong}, {Mackey},
  {Marino}, {Monty}, {Nordlander}, {Norris}, {Asplund}, {Bessell}, {Casey},
  {Frebel}, {Lind}, {Murphy}, {Schmidt}, {Gao}, {Xylakis-Dornbusch}, {Amarsi},
  \& {Milone}}]{cordoni21}
{Cordoni}, G., {Da Costa}, G.~S., {Yong}, D., {et~al.} 2021, \mnras, 503, 2539

\bibitem[{{Dillamore} {et~al.}(2023{\natexlab{a}}){Dillamore}, {Belokurov},
  {Evans}, \& {Davies}}]{dillamore_bar}
{Dillamore}, A.~M., {Belokurov}, V., {Evans}, N.~W., \& {Davies}, E.~Y.
  2023{\natexlab{a}}, \mnras, 524, 3596

\bibitem[{{Dillamore} {et~al.}(2023{\natexlab{b}}){Dillamore}, {Belokurov},
  {Kravtsov}, \& {Font}}]{dillamore23}
{Dillamore}, A.~M., {Belokurov}, V., {Kravtsov}, A., \& {Font}, A.~S.
  2023{\natexlab{b}}, arXiv e-prints, arXiv:2309.08658

\bibitem[{{Dodd} {et~al.}(2023){Dodd}, {Callingham}, {Helmi}, {Matsuno},
  {Ruiz-Lara}, {Balbinot}, \& {L{\"o}vdal}}]{dodd23}
{Dodd}, E., {Callingham}, T.~M., {Helmi}, A., {et~al.} 2023, \aap, 670, L2

\bibitem[{{Dovgal} {et~al.}(2023){Dovgal}, {Venn}, {Sestito}, {Hayes},
  {McConnachie}, {Navarro}, {Placco}, {Starkenburg}, {Martin}, {Pazder},
  {Chiboucas}, {Deibert}, {Gamen}, {Heo}, {Kalari}, {Martioli}, {Xu}, {Diaz},
  {Gomez-Jiminez}, {Henderson}, {Prado}, {Quiroz}, {Robertson}, {Ruiz-Carmona},
  {Simpson}, {Urrutia}, {Waller}, {Berg}, {Burley}, {Hartman}, {Ireland},
  {Margheim}, {Perez}, \& {Thomas-Osip}}]{dovgal23}
{Dovgal}, A., {Venn}, K.~A., {Sestito}, F., {et~al.} 2023, arXiv e-prints,
  arXiv:2310.03075

\bibitem[{{El-Badry} {et~al.}(2018){El-Badry}, {Bland-Hawthorn}, {Wetzel},
  {Quataert}, {Weisz}, {Boylan-Kolchin}, {Hopkins}, {Faucher-Gigu{\`e}re},
  {Kere{\v{s}}}, \& {Garrison-Kimmel}}]{elbadry18}
{El-Badry}, K., {Bland-Hawthorn}, J., {Wetzel}, A., {et~al.} 2018, \mnras, 480,
  652

\bibitem[{{Gaia Collaboration} {et~al.}(2023){Gaia Collaboration}, {Vallenari},
  {Brown}, {Prusti}, {de Bruijne}, {Arenou}, {Babusiaux}, {Biermann},
  {Creevey}, {Ducourant}, {Evans}, {Eyer}, {Guerra}, {Hutton}, {Jordi},
  {Klioner}, {Lammers}, {Lindegren}, {Luri}, {Mignard}, {Panem}, {Pourbaix},
  {Randich}, {Sartoretti}, {Soubiran}, {Tanga}, {Walton}, {Bailer-Jones},
  {Bastian}, {Drimmel}, {Jansen}, {Katz}, {Lattanzi}, {van Leeuwen}, {Bakker},
  {Cacciari}, {Casta{\~n}eda}, {De Angeli}, {Fabricius}, {Fouesneau},
  {Fr{\'e}mat}, {Galluccio}, {Guerrier}, {Heiter}, {Masana}, {Messineo},
  {Mowlavi}, {Nicolas}, {Nienartowicz}, {Pailler}, {Panuzzo}, {Riclet}, {Roux},
  {Seabroke}, {Sordo}, {Th{\'e}venin}, {Gracia-Abril}, {Portell}, {Teyssier},
  {Altmann}, {Andrae}, {Audard}, {Bellas-Velidis}, {Benson}, {Berthier},
  {Blomme}, {Burgess}, {Busonero}, {Busso}, {C{\'a}novas}, {Carry}, {Cellino},
  {Cheek}, {Clementini}, {Damerdji}, {Davidson}, {de Teodoro}, {Nu{\~n}ez
  Campos}, {Delchambre}, {Dell'Oro}, {Esquej}, {Fern{\'a}ndez-Hern{\'a}ndez},
  {Fraile}, {Garabato}, {Garc{\'\i}a-Lario}, {Gosset}, {Haigron}, {Halbwachs},
  {Hambly}, {Harrison}, {Hern{\'a}ndez}, {Hestroffer}, {Hodgkin}, {Holl},
  {Jan{\ss}en}, {Jevardat de Fombelle}, {Jordan}, {Krone-Martins}, {Lanzafame},
  {L{\"o}ffler}, {Marchal}, {Marrese}, {Moitinho}, {Muinonen}, {Osborne},
  {Pancino}, {Pauwels}, {Recio-Blanco}, {Reyl{\'e}}, {Riello}, {Rimoldini},
  {Roegiers}, {Rybizki}, {Sarro}, {Siopis}, {Smith}, {Sozzetti}, {Utrilla},
  {van Leeuwen}, {Abbas}, {{\'A}brah{\'a}m}, {Abreu Aramburu}, {Aerts},
  {Aguado}, {Ajaj}, {Aldea-Montero}, {Altavilla}, {{\'A}lvarez}, {Alves},
  {Anders}, {Anderson}, {Anglada Varela}, {Antoja}, {Baines}, {Baker},
  {Balaguer-N{\'u}{\~n}ez}, {Balbinot}, {Balog}, {Barache}, {Barbato},
  {Barros}, {Barstow}, {Bartolom{\'e}}, {Bassilana}, {Bauchet}, {Becciani},
  {Bellazzini}, {Berihuete}, {Bernet}, {Bertone}, {Bianchi}, {Binnenfeld},
  {Blanco-Cuaresma}, {Blazere}, {Boch}, {Bombrun}, {Bossini}, {Bouquillon},
  {Bragaglia}, {Bramante}, {Breedt}, {Bressan}, {Brouillet}, {Brugaletta},
  {Bucciarelli}, {Burlacu}, {Butkevich}, {Buzzi}, {Caffau}, {Cancelliere},
  {Cantat-Gaudin}, {Carballo}, {Carlucci}, {Carnerero}, {Carrasco},
  {Casamiquela}, {Castellani}, {Castro-Ginard}, {Chaoul}, {Charlot}, {Chemin},
  {Chiaramida}, {Chiavassa}, {Chornay}, {Comoretto}, {Contursi}, {Cooper},
  {Cornez}, {Cowell}, {Crifo}, {Cropper}, {Crosta}, {Crowley}, {Dafonte},
  {Dapergolas}, {David}, {David}, {de Laverny}, {De Luise}, {De March}, {De
  Ridder}, {de Souza}, {de Torres}, {del Peloso}, {del Pozo}, {Delbo},
  {Delgado}, {Delisle}, {Demouchy}, {Dharmawardena}, {Di Matteo}, {Diakite},
  {Diener}, {Distefano}, {Dolding}, {Edvardsson}, {Enke}, {Fabre}, {Fabrizio},
  {Faigler}, {Fedorets}, {Fernique}, {Fienga}, {Figueras}, {Fournier},
  {Fouron}, {Fragkoudi}, {Gai}, {Garcia-Gutierrez}, {Garcia-Reinaldos},
  {Garc{\'\i}a-Torres}, {Garofalo}, {Gavel}, {Gavras}, {Gerlach}, {Geyer},
  {Giacobbe}, {Gilmore}, {Girona}, {Giuffrida}, {Gomel}, {Gomez},
  {Gonz{\'a}lez-N{\'u}{\~n}ez}, {Gonz{\'a}lez-Santamar{\'\i}a},
  {Gonz{\'a}lez-Vidal}, {Granvik}, {Guillout}, {Guiraud},
  {Guti{\'e}rrez-S{\'a}nchez}, {Guy}, {Hatzidimitriou}, {Hauser}, {Haywood},
  {Helmer}, {Helmi}, {Sarmiento}, {Hidalgo}, {Hilger}, {H{\l}adczuk}, {Hobbs},
  {Holland}, {Huckle}, {Jardine}, {Jasniewicz}, {Jean-Antoine Piccolo},
  {Jim{\'e}nez-Arranz}, {Jorissen}, {Juaristi Campillo}, {Julbe}, {Karbevska},
  {Kervella}, {Khanna}, {Kontizas}, {Kordopatis}, {Korn}, {K{\'o}sp{\'a}l},
  {Kostrzewa-Rutkowska}, {Kruszy{\'n}ska}, {Kun}, {Laizeau}, {Lambert},
  {Lanza}, {Lasne}, {Le Campion}, {Lebreton}, {Lebzelter}, {Leccia}, {Leclerc},
  {Lecoeur-Taibi}, {Liao}, {Licata}, {Lindstr{\o}m}, {Lister}, {Livanou},
  {Lobel}, {Lorca}, {Loup}, {Madrero Pardo}, {Magdaleno Romeo}, {Managau},
  {Mann}, {Manteiga}, {Marchant}, {Marconi}, {Marcos}, {Marcos Santos},
  {Mar{\'\i}n Pina}, {Marinoni}, {Marocco}, {Marshall}, {Martin Polo},
  {Mart{\'\i}n-Fleitas}, {Marton}, {Mary}, {Masip}, {Massari},
  {Mastrobuono-Battisti}, {Mazeh}, {McMillan}, {Messina}, {Michalik}, {Millar},
  {Mints}, {Molina}, {Molinaro}, {Moln{\'a}r}, {Monari}, {Mongui{\'o}},
  {Montegriffo}, {Montero}, {Mor}, {Mora}, {Morbidelli}, {Morel}, {Morris},
  {Muraveva}, {Murphy}, {Musella}, {Nagy}, {Noval}, {Oca{\~n}a}, {Ogden},
  {Ordenovic}, {Osinde}, {Pagani}, {Pagano}, {Palaversa}, {Palicio},
  {Pallas-Quintela}, {Panahi}, {Payne-Wardenaar}, {Pe{\~n}alosa Esteller},
  {Penttil{\"a}}, {Pichon}, {Piersimoni}, {Pineau}, {Plachy}, {Plum}, {Poggio},
  {Pr{\v{s}}a}, {Pulone}, {Racero}, {Ragaini}, {Rainer}, {Raiteri}, {Rambaux},
  {Ramos}, {Ramos-Lerate}, {Re Fiorentin}, {Regibo}, {Richards}, {Rios Diaz},
  {Ripepi}, {Riva}, {Rix}, {Rixon}, {Robichon}, {Robin}, {Robin}, {Roelens},
  {Rogues}, {Rohrbasser}, {Romero-G{\'o}mez}, {Rowell}, {Royer}, {Ruz Mieres},
  {Rybicki}, {Sadowski}, {S{\'a}ez N{\'u}{\~n}ez}, {Sagrist{\`a} Sell{\'e}s},
  {Sahlmann}, {Salguero}, {Samaras}, {Sanchez Gimenez}, {Sanna},
  {Santove{\~n}a}, {Sarasso}, {Schultheis}, {Sciacca}, {Segol}, {Segovia},
  {S{\'e}gransan}, {Semeux}, {Shahaf}, {Siddiqui}, {Siebert}, {Siltala},
  {Silvelo}, {Slezak}, {Slezak}, {Smart}, {Snaith}, {Solano}, {Solitro},
  {Souami}, {Souchay}, {Spagna}, {Spina}, {Spoto}, {Steele},
  {Steidelm{\"u}ller}, {Stephenson}, {S{\"u}veges}, {Surdej}, {Szabados},
  {Szegedi-Elek}, {Taris}, {Taylor}, {Teixeira}, {Tolomei}, {Tonello}, {Torra},
  {Torra}, {Torralba Elipe}, {Trabucchi}, {Tsounis}, {Turon}, {Ulla}, {Unger},
  {Vaillant}, {van Dillen}, {van Reeven}, {Vanel}, {Vecchiato}, {Viala},
  {Vicente}, {Voutsinas}, {Weiler}, {Wevers}, {Wyrzykowski}, {Yoldas}, {Yvard},
  {Zhao}, {Zorec}, {Zucker}, \& {Zwitter}}]{dr3_general}
{Gaia Collaboration}, {Vallenari}, A., {Brown}, A.~G.~A., {et~al.} 2023, \aap,
  674, A1

\bibitem[{{Grand} {et~al.}(2017){Grand}, {G{\'o}mez}, {Marinacci}, {Pakmor},
  {Springel}, {Campbell}, {Frenk}, {Jenkins}, \& {White}}]{grand17}
{Grand}, R. J.~J., {G{\'o}mez}, F.~A., {Marinacci}, F., {et~al.} 2017, \mnras,
  467, 179

\bibitem[{{Hawkins} {et~al.}(2015){Hawkins}, {Jofr{\'e}}, {Masseron}, \&
  {Gilmore}}]{hawkins15}
{Hawkins}, K., {Jofr{\'e}}, P., {Masseron}, T., \& {Gilmore}, G. 2015, \mnras,
  453, 758

\bibitem[{{Helmi} {et~al.}(2018){Helmi}, {Babusiaux}, {Koppelman}, {Massari},
  {Veljanoski}, \& {Brown}}]{helmi18}
{Helmi}, A., {Babusiaux}, C., {Koppelman}, H.~H., {et~al.} 2018, \nat, 563, 85

\bibitem[{{Hong} {et~al.}(2023){Hong}, {Beers}, {Lee}, {Huang}, {Hirai},
  {Cabrera Garcia}, {Shank}, {Xu}, {Mardini}, {Catapano}, {Zhao}, {Fan},
  {Zheng}, {Wang}, {Tan}, {Zhao}, \& {Li}}]{hong23}
{Hong}, J., {Beers}, T.~C., {Lee}, Y.~S., {et~al.} 2023, arXiv e-prints,
  arXiv:2311.02297

\bibitem[{{Hopkins} {et~al.}(2018){Hopkins}, {Wetzel}, {Kere{\v{s}}},
  {Faucher-Gigu{\`e}re}, {Quataert}, {Boylan-Kolchin}, {Murray}, {Hayward},
  {Garrison-Kimmel}, {Hummels}, {Feldmann}, {Torrey}, {Ma},
  {Angl{\'e}s-Alc{\'a}zar}, {Su}, {Orr}, {Schmitz}, {Escala}, {Sanderson},
  {Grudi{\'c}}, {Hafen}, {Kim}, {Fitts}, {Bullock}, {Wheeler}, {Chan},
  {Elbert}, \& {Narayanan}}]{hopkins18}
{Hopkins}, P.~F., {Wetzel}, A., {Kere{\v{s}}}, D., {et~al.} 2018, \mnras, 480,
  800

\bibitem[{{Ibata} {et~al.}(2021){Ibata}, {Malhan}, {Martin}, {Aubert},
  {Famaey}, {Bianchini}, {Monari}, {Siebert}, {Thomas}, {Bellazzini},
  {Bonifacio}, {Caffau}, \& {Renaud}}]{iba21}
{Ibata}, R., {Malhan}, K., {Martin}, N., {et~al.} 2021, \apj, 914, 123

\bibitem[{{Ibata} {et~al.}(2019){Ibata}, {Malhan}, \& {Martin}}]{iba19}
{Ibata}, R.~A., {Malhan}, K., \& {Martin}, N.~F. 2019, \apj, 872, 152

\bibitem[{{Ivezi{\'c}} {et~al.}(2008){Ivezi{\'c}}, {Sesar}, {Juri{\'c}},
  {Bond}, {Dalcanton}, {Rockosi}, {Yanny}, {Newberg}, {Beers}, {Allende
  Prieto}, {Wilhelm}, {Lee}, {Sivarani}, {Norris}, {Bailer-Jones}, {Re
  Fiorentin}, {Schlegel}, {Uomoto}, {Lupton}, {Knapp}, {Gunn}, {Covey}, {Allyn
  Smith}, {Miknaitis}, {Doi}, {Tanaka}, {Fukugita}, {Kent}, {Finkbeiner},
  {Munn}, {Pier}, {Quinn}, {Hawley}, {Anderson}, {Kiuchi}, {Chen}, {Bushong},
  {Sohi}, {Haggard}, {Kimball}, {Barentine}, {Brewington}, {Harvanek},
  {Kleinman}, {Krzesinski}, {Long}, {Nitta}, {Snedden}, {Lee}, {Harris},
  {Brinkmann}, {Schneider}, \& {York}}]{ivezic08}
{Ivezi{\'c}}, {\v{Z}}., {Sesar}, B., {Juri{\'c}}, M., {et~al.} 2008, \apj, 684,
  287

\bibitem[{{Katz} {et~al.}(2022){Katz}, {Sartoretti}, {Guerrier}, {Panuzzo},
  {Seabroke}, {Th{\'e}venin}, {Cropper}, {Benson}, {Blomme}, {Haigron},
  {Marchal}, {Smith}, {Baker}, {Chemin}, {Damerdji}, {David}, {Dolding},
  {Fr{\'e}mat}, {Gosset}, {Jan{\ss}en}, {Jasniewicz}, {Lobel}, {Plum},
  {Samaras}, {Snaith}, {Soubiran}, {Vanel}, {Zwitter}, {Antoja}, {Arenou},
  {Babusiaux}, {Brouillet}, {Caffau}, {Di Matteo}, {Fabre}, {Fabricius},
  {Frakgoudi}, {Haywood}, {Huckle}, {Hottier}, {Lasne}, {Leclerc},
  {Mastrobuono-Battisti}, {Royer}, {Teyssier}, {Zorec}, {Crifo}, {Jean-Antoine
  Piccolo}, {Turon}, \& {Viala}}]{dr3_rv}
{Katz}, D., {Sartoretti}, P., {Guerrier}, A., {et~al.} 2022, arXiv e-prints,
  arXiv:2206.05902

\bibitem[{{Koppelman} {et~al.}(2019{\natexlab{a}}){Koppelman}, {Helmi},
  {Massari}, {Price-Whelan}, \& {Starkenburg}}]{koppelman19b}
{Koppelman}, H.~H., {Helmi}, A., {Massari}, D., {Price-Whelan}, A.~M., \&
  {Starkenburg}, T.~K. 2019{\natexlab{a}}, \aap, 631, L9

\bibitem[{{Koppelman} {et~al.}(2019{\natexlab{b}}){Koppelman}, {Helmi},
  {Massari}, {Roelenga}, \& {Bastian}}]{koppelman19}
{Koppelman}, H.~H., {Helmi}, A., {Massari}, D., {Roelenga}, S., \& {Bastian},
  U. 2019{\natexlab{b}}, \aap, 625, A5

\bibitem[{{Leung} \& {Bovy}(2019)}]{astronn}
{Leung}, H.~W. \& {Bovy}, J. 2019, \mnras, 489, 2079

\bibitem[{{Li} \& {Zhao}(2017)}]{li_zhao17}
{Li}, C. \& {Zhao}, G. 2017, \apj, 850, 25

\bibitem[{{Malhan} {et~al.}(2021){Malhan}, {Yuan}, {Ibata}, {Arentsen},
  {Bellazzini}, \& {Martin}}]{lms-1}
{Malhan}, K., {Yuan}, Z., {Ibata}, R.~A., {et~al.} 2021, \apj, 920, 51

\bibitem[{{Martin} {et~al.}(2022){Martin}, {Ibata}, {Starkenburg}, {Yuan},
  {Malhan}, {Bellazzini}, {Viswanathan}, {Aguado}, {Arentsen}, {Bonifacio},
  {Carlberg}, {Gonz{\'a}lez Hern{\'a}ndez}, {Hill}, {Jablonka}, {Kordopatis},
  {Lardo}, {McConnachie}, {Navarro}, {S{\'a}nchez-Janssen}, {Sestito},
  {Thomas}, {Venn}, {Vitali}, \& {Voggel}}]{pristine_streams}
{Martin}, N.~F., {Ibata}, R.~A., {Starkenburg}, E., {et~al.} 2022, \mnras, 516,
  5331

\bibitem[{{Martin} {et~al.}(2023){Martin}, {Starkenburg}, {Yuan}, {Fouesneau},
  {Arentsen}, {De Angeli}, {Gran}, {Montelius}, {Andrae}, {Bellazzini},
  {Montegriffo}, {Esselink}, {Zhang}, {Venn}, {Viswanathan}, {Aguado},
  {Battaglia}, {Bayer}, {Bonifacio}, {Caffau}, {C{\^o}t{\'e}}, {Carlberg},
  {Fabbro}, {Fern{\'a}ndez Alvar}, {Gonz{\'a}lez Hern{\'a}ndez}, {Gonz{\'a}lez
  Rivera de La Vernhe}, {Hill}, {Ibata}, {Jablonka}, {Kordopatis}, {Lardo},
  {McConnachie}, {Navarrete}, {Navarro}, {Recio-Blanco}, {S{\'a}nchez Janssen},
  {Sestito}, {Thomas}, {Vitali}, \& {Youakim}}]{martin23}
{Martin}, N.~F., {Starkenburg}, E., {Yuan}, Z., {et~al.} 2023, arXiv e-prints,
  arXiv:2308.01344

\bibitem[{{McMillan}(2017)}]{mcmillan17}
{McMillan}, P.~J. 2017, \mnras, 465, 76

\bibitem[{{Miranda} {et~al.}(2016){Miranda}, {Pilkington}, {Gibson}, {Brook},
  {S{\'a}nchez-Bl{\'a}zquez}, {Minchev}, {Few}, {Smith},
  {Dom{\'\i}nguez-Tenreiro}, {Obreja}, {Bailin}, \& {Stinson}}]{miranda16}
{Miranda}, M.~S., {Pilkington}, K., {Gibson}, B.~K., {et~al.} 2016, \aap, 587,
  A10

\bibitem[{{Myeong} {et~al.}(2019){Myeong}, {Vasiliev}, {Iorio}, {Evans}, \&
  {Belokurov}}]{myeong19}
{Myeong}, G.~C., {Vasiliev}, E., {Iorio}, G., {Evans}, N.~W., \& {Belokurov},
  V. 2019, \mnras, 488, 1235

\bibitem[{{Naidu} {et~al.}(2020){Naidu}, {Conroy}, {Bonaca}, {Johnson}, {Ting},
  {Caldwell}, {Zaritsky}, \& {Cargile}}]{naidu20}
{Naidu}, R.~P., {Conroy}, C., {Bonaca}, A., {et~al.} 2020, \apj, 901, 48

\bibitem[{{Necib} {et~al.}(2020){Necib}, {Ostdiek}, {Lisanti}, {Cohen},
  {Freytsis}, {Garrison-Kimmel}, {Hopkins}, {Wetzel}, \& {Sanderson}}]{nyx}
{Necib}, L., {Ostdiek}, B., {Lisanti}, M., {et~al.} 2020, Nature Astronomy, 4,
  1078

\bibitem[{{Pillepich} {et~al.}(2018){Pillepich}, {Nelson}, {Hernquist},
  {Springel}, {Pakmor}, {Torrey}, {Weinberger}, {Genel}, {Naiman}, {Marinacci},
  \& {Vogelsberger}}]{pillepich18}
{Pillepich}, A., {Nelson}, D., {Hernquist}, L., {et~al.} 2018, \mnras, 475, 648

\bibitem[{{Pinna} {et~al.}(2023){Pinna}, {Walo-Mart{\'\i}n}, {Grand}, {Martig},
  {Fragkoudi}, {G{\'o}mez}, {Marinacci}, \& {Pakmor}}]{pinna23}
{Pinna}, F., {Walo-Mart{\'\i}n}, D., {Grand}, R. J.~J., {et~al.} 2023, arXiv
  e-prints, arXiv:2311.13700

\bibitem[{{Re Fiorentin} {et~al.}(2021){Re Fiorentin}, {Spagna}, {Lattanzi}, \&
  {Cignoni}}]{refiorentin21}
{Re Fiorentin}, P., {Spagna}, A., {Lattanzi}, M.~G., \& {Cignoni}, M. 2021,
  \apjl, 907, L16

\bibitem[{{Rix} {et~al.}(2022){Rix}, {Chandra}, {Andrae}, {Price-Whelan},
  {Weinberg}, {Conroy}, {Fouesneau}, {Hogg}, {De Angeli}, {Naidu}, {Xiang}, \&
  {Ruz-Mieres}}]{rix22}
{Rix}, H.-W., {Chandra}, V., {Andrae}, R., {et~al.} 2022, \apj, 941, 45

\bibitem[{{Ruchti} {et~al.}(2010){Ruchti}, {Fulbright}, {Wyse}, {Gilmore},
  {Bienaym{\'e}}, {Binney}, {Bland-Hawthorn}, {Campbell}, {Freeman}, {Gibson},
  {Grebel}, {Helmi}, {Munari}, {Navarro}, {Parker}, {Reid}, {Seabroke},
  {Siebert}, {Siviero}, {Steinmetz}, {Watson}, {Williams}, \&
  {Zwitter}}]{ruchti10}
{Ruchti}, G.~R., {Fulbright}, J.~P., {Wyse}, R.~F.~G., {et~al.} 2010, \apjl,
  721, L92

\bibitem[{{Santistevan} {et~al.}(2021){Santistevan}, {Wetzel}, {Sanderson},
  {El-Badry}, {Samuel}, \& {Faucher-Gigu{\`e}re}}]{santistevan21}
{Santistevan}, I.~B., {Wetzel}, A., {Sanderson}, R.~E., {et~al.} 2021, \mnras,
  505, 921

\bibitem[{{Semenov} {et~al.}(2023{\natexlab{a}}){Semenov}, {Conroy}, {Chandra},
  {Hernquist}, \& {Nelson}}]{semenov23a}
{Semenov}, V.~A., {Conroy}, C., {Chandra}, V., {Hernquist}, L., \& {Nelson}, D.
  2023{\natexlab{a}}, arXiv e-prints, arXiv:2306.09398

\bibitem[{{Semenov} {et~al.}(2023{\natexlab{b}}){Semenov}, {Conroy}, {Chandra},
  {Hernquist}, \& {Nelson}}]{semenov23b}
{Semenov}, V.~A., {Conroy}, C., {Chandra}, V., {Hernquist}, L., \& {Nelson}, D.
  2023{\natexlab{b}}, arXiv e-prints, arXiv:2306.13125

\bibitem[{{Sestito} {et~al.}(2021){Sestito}, {Buck}, {Starkenburg}, {Martin},
  {Navarro}, {Venn}, {Obreja}, {Jablonka}, \& {Macci{\`o}}}]{sestito21}
{Sestito}, F., {Buck}, T., {Starkenburg}, E., {et~al.} 2021, \mnras, 500, 3750

\bibitem[{{Sestito} {et~al.}(2019){Sestito}, {Longeard}, {Martin},
  {Starkenburg}, {Fouesneau}, {Gonz{\'a}lez Hern{\'a}ndez}, {Arentsen},
  {Ibata}, {Aguado}, {Carlberg}, {Jablonka}, {Navarro}, {Tolstoy}, \&
  {Venn}}]{sestito19}
{Sestito}, F., {Longeard}, N., {Martin}, N.~F., {et~al.} 2019, \mnras, 484,
  2166

\bibitem[{{Sestito} {et~al.}(2020){Sestito}, {Martin}, {Starkenburg},
  {Arentsen}, {Ibata}, {Longeard}, {Kielty}, {Youakim}, {Venn}, {Aguado},
  {Carlberg}, {Gonz{\'a}lez Hern{\'a}ndez}, {Hill}, {Jablonka}, {Kordopatis},
  {Malhan}, {Navarro}, {S{\'a}nchez-Janssen}, {Thomas}, {Tolstoy}, {Wilson},
  {Palicio}, {Bialek}, {Garcia-Dias}, {Lucchesi}, {North}, {Osorio}, {Patrick},
  \& {Peralta de Arriba}}]{sestito20}
{Sestito}, F., {Martin}, N.~F., {Starkenburg}, E., {et~al.} 2020, \mnras, 497,
  L7

\bibitem[{{Sotillo-Ramos} {et~al.}(2023){Sotillo-Ramos}, {Bergemann}, {Friske},
  \& {Pillepich}}]{sotillo23}
{Sotillo-Ramos}, D., {Bergemann}, M., {Friske}, J. K.~S., \& {Pillepich}, A.
  2023, \mnras [\eprint[arXiv]{2307.14421}]

\bibitem[{{Taylor}(2005)}]{Taylor2005}
{Taylor}, M.~B. 2005, in Astronomical Society of the Pacific Conference Series,
  Vol. 347, Astronomical Data Analysis Software and Systems XIV, ed.
  P.~{Shopbell}, M.~{Britton}, \& R.~{Ebert}, 29

\bibitem[{{Vasiliev}(2019)}]{vasiliev19}
{Vasiliev}, E. 2019, \mnras, 482, 1525

\bibitem[{{Xiang} \& {Rix}(2022)}]{xiang_rix22}
{Xiang}, M. \& {Rix}, H.-W. 2022, \nat, 603, 599

\bibitem[{{Xu} {et~al.}(2022){Xu}, {Yuan}, {Niu}, {Yang}, {Beers}, \&
  {Huang}}]{xu22}
{Xu}, S., {Yuan}, H., {Niu}, Z., {et~al.} 2022, \apjs, 258, 44

\bibitem[{{Yang} {et~al.}(2022){Yang}, {Yuan}, {Xiang}, {Duan}, {Huang}, {Liu},
  {Beers}, {Galarza}, {Daflon}, {Fern{\'a}ndez-Ontiveros}, {Cenarro},
  {Crist{\'o}bal-Hornillos}, {Hern{\'a}ndez-Monteagudo}, {L{\'o}pez-Sanjuan},
  {Mar{\'\i}n-Franch}, {Moles}, {Varela}, {Rami{\'o}}, {Alcaniz}, {Dupke},
  {Ederoclite}, {Sodr{\'e}}, \& {Angulo}}]{yang22}
{Yang}, L., {Yuan}, H., {Xiang}, M., {et~al.} 2022, \aap, 659, A181

\bibitem[{{Yuan} {et~al.}(2023){Yuan}, {Li}, {Martin}, {Monari}, {Famaey},
  {Siebert}, {Ardern-Arentsen}, {Sestito}, {Thomas}, {Hill}, {Ibata},
  {Kordopatis}, {Starkenburg}, \& {Viswanathan}}]{yuan23}
{Yuan}, Z., {Li}, C., {Martin}, N.~F., {et~al.} 2023, arXiv e-prints,
  arXiv:2311.08464

\bibitem[{{Zhang} {et~al.}(2023){Zhang}, {Ardern-Arentsen}, \&
  {Belokurov}}]{zhang23}
{Zhang}, H., {Ardern-Arentsen}, A., \& {Belokurov}, V. 2023, arXiv e-prints,
  arXiv:2311.09294

\end{thebibliography}





\end{document}